\documentstyle[12pt,epsf,floatfig]{article}
\def\PrePrintE#1{\mbox{hep-ex/#1}}
\def\PLB{{\em Phys. Lett.}  B}
\def\PrePrintP#1{\mbox{hep-ph/#1}}
\def\stop{\tilde{t}}

 \newcommand{\EEMASS               }{  163 \pm 3}

\newcommand{\twofig}[8]
{\begin{figure}[tb]
\vspace{1.5cm}
 \begin{minipage}[t]{0.48\linewidth}
   \epsfxsize \linewidth
   \epsffile[10 128 550 675]{#1}
   \vspace{#4}
   \vspace{-0.5cm}
   \caption{#2}
   \label{#3}
 \end{minipage}
 \hfil
 \begin{minipage}[t]{0.480\linewidth}
   \epsfxsize \linewidth
   \epsffile[10 128 550 675]{#5}
   \vspace{#8}
   \vspace{-0.5cm}
   \caption{#6}
   \label{#7}
 \end{minipage}
\vspace{-1.5cm}
\end{figure}
}

\newcommand{\dtfloat}[4]
{
\begin{floatingfigure}{0.48\linewidth}
\vspace*{0.5cm}
\hspace*{-0.5cm}
\epsfxsize #4\linewidth
\epsffile[30 128 520 620]{#1}
\vspace*{-0.3cm}
\caption{#2}
\label{#3}
\end{floatingfigure}
}

\newcommand{\mett}{\mbox{${\rm \not\! E}_{\rm T}$}}
\newcommand{\eeggmett}{ee\gamma\gamma\mett}
\def\gravitino{\tilde{G}}
\newcommand{\NONE}{\mbox{$N_1$}}
\newcommand{\Etggl}{\mbox{${\rm E}_{\rm T}^{\gamma}>12$~GeV}}
\newcommand{\Etggh}{\mbox{${\rm E}_{\rm T}^{\gamma}>25$~GeV}}
\def \Et {{\rm E}_{\rm T}}
\def\Z{{ Z^0}}
\newcommand{\mettgmh}{\mbox{${\rm \not\! E}_{\rm T}>35$~GeV}}
\newcommand{\mettgl}{\mbox{${\rm \not\! E}_{\rm T}>25$~GeV}}
\newcommand{\mettsm}{\mbox{\scriptsize ${\rm \not\! E}_{\rm T}$}}
\newcommand{\ptt}{${\rm P}_{\rm T}$}

\newcommand{\zoee}{\mbox{$Z^0 \rightarrow e^+e^-$}}
\newcommand{\scinotn}[2]{\mbox{${#1}\times 10^{#2}$}}
\newcommand{\etal}{{\em et al.}}
\newcommand{\NTGNO}{\mbox{$\NTWO \rightarrow \gamma \NONE$}}
\newcommand{\NTWO}{\mbox{$N_2$}}
\def\roots{{\sqrt s}}

 \newcommand{\NMETEXPLOW           }{\mbox{$     0.5\pm      0.1$}}
 \newcommand{\NMETEXPHIGH          }{\mbox{$     0.5\pm      0.1$}}
 \newcommand{\NMETLOW              }{    1}
 \newcommand{\NMETHIGH             }{    2}

 \newcommand{\NIVJETEXPLOW         }{\mbox{$     1.6\pm      0.4$}}
 \newcommand{\NIVJETLOW            }{    2}
 \newcommand{\NIIIJETEXPHIGH       }{\mbox{$     1.7\pm      1.5$}}
 \newcommand{\NIIIJETHIGH          }{    0}
 \newcommand{\NCENTEORMU           }{    3}

 \newcommand{\MUMUGGMASS           }{   92 \pm 1}
 \newcommand{\EGGMASS              }{   91 \pm 2}

 \newcommand{\NCENTEORMUEXP        }{\mbox{$     0.3\pm      0.1$}}

 \newcommand{\NTAUEXPLOW           }{\mbox{$     0.2\pm      0.1$}}
 
 \newcommand{\NCENTTAU             }{    1}

 \newcommand{\NBEXPLOW             }{\mbox{$     1.3\pm      0.7$}}
 \newcommand{\NBTAG                }{    2}

 \newcommand{\NADDGAMMAEXP         }{\mbox{$     0.1\pm      0.1$}}
 \newcommand{\NADDGAMMA            }{    0}

 \newcommand{\EEGGPT               }{   48 \pm 2}

 \newcommand{\EEFAKEGGMETTOTRATE   }{\scinotn{(6\pm 6)}{ -8}}
 
\newcommand{\CONE}{\mbox{$C_1$}}

\def\Journal#1#2#3#4{{#1} {\bf #2}, #3 (#4)}
\def\PRL{\em Phys. Rev. Lett.}
\def\PRD{{\em Phys. Rev.} D}

\def\PrePrintE#1{\mbox{hep-ex/#1}}
\def\PrePrintP#1{\mbox{hep-ph/#1}}

\begin{document}
\initfloatingfigs

\topskip 2cm 
\begin{titlepage}

%\begin{flushright}
%    CDF/PUB/EXOTIC/PUBLIC/4612\\
%    Version 1.0\\
%    \today\\
%\end{flushright}

\begin{center}
{\large\bf CDF SEARCHES FOR NEW PHENOMENA} \\
\vspace{2.5cm}
{\large Dave Toback} \\
\vspace{.5cm}
{\sl University of Chicago\\ for  the CDF
Collaboration }\\
\vspace{2.5cm}
\vfil

\begin{abstract} 

\begin{sloppypar}
We present results of recent searches for new physics 
using the CDF detector at the Fermilab Tevatron.  Presented are
searches for \mbox{Higgs $\rightarrow \gamma\gamma$,} as well as more 
general searches in 
$\gamma\gamma + X$ where $X$ is 
$\mett$, jets, charged leptons, $b$-quarks or extra photons.  The CDF 
$\eeggmett$ candidate event is discussed along with estimates of the expected
rates in the Standard Model. Other searches for SUSY, Higgs and
Technicolor look for particles which decay to vector bosons and $b$-quarks.
\end{sloppypar}
\end{abstract}

\end{center}
\end{titlepage}

%******************************************************************************

\section{Introduction}

Many possible extensions to the Standard Model include final state
signatures involving
vector bosons. For example, events which can decay to final state photons have
recently received a great deal of attention.  We present
resent results from data collected by the CDF detector~\cite{CDF Detector} 
during the 1992-1995
collider run of the Fermilab Tevatron.

\begin{sloppypar}
We begin with a search for new heavy particles which decay into two photons 
e.g. \mbox{Higgs $\rightarrow \gamma\gamma$~\cite{Higgs}.} Other searches in
two photon events look for 
missing E$_{\rm T}$, jets,
charged leptons, $b$-quarks and/or extra photons and provide sensitivity
to Supersymmetry and other 
theoretical models~\cite{Gravitino Reference, Higgsino LSP, Non-SUSY}.
The CDF $\eeggmett$ candidate event~\cite{GG PRL} turns up naturally 
in these 
searches and we present the 
results of a detailed study of the event. 
The expected number of
events with the $\eeggmett$ signature 
from Standard Model production are summarized.  A number of models which
predict $\eeggmett$ events also predict events in the inclusive $\gamma\gamma
+\mett$ and $\gamma\gamma + $jets 
data and  limits are presented on these models. 
We conclude with searches for Supersymmetry, associated 
Higgs production and Technicolor by looking
for
new particles which decay to final states which contain a vector boson
($\gamma, W$ or $Z^0$) and a 
$b$-quark~\cite{Higgs, Higgsino LSP, Techni Omega, Techni Rho}. 
\end{sloppypar}

%Events with a photons and a $b$-quark can provide a new way to search for hints
%of Supersymmetry~\cite{Higgsino LSP}c and 
%Technicolor~\cite{Techni Omega}.  Other models of
%Technicolor models also predict final state $b$-quarks produced in association
%with heavy vector bosons. These same datasets can be used to search for
%associated Higgs production~\cite{Higgs}.

% ******************************************************************************

\section{High Mass $X\rightarrow \gamma\gamma$ Search}

%Two Photon Events}

Many models of new physics predict events with two photons in the final state.
We begin with a search for new heavy particles
 which decay via \mbox{$X \rightarrow \gamma\gamma$}. 
The invariant mass of isolated diphoton candidates
with ${\rm E_T^{\gamma}>22}$~GeV and $|\eta|<1.0$ in 100$\pm 7$~pb$^{-1}$ of 
data is
shown in Figure~\ref{gg_xsec} along with the
Standard Model predictions.
There is no evidence for an excess or resonance at
high mass.  Using a model of direct Higgs production in
 which the Higgs decays via 
\mbox{$H \rightarrow \gamma\gamma$}~\cite{Higgs},
an experimental limit is
set as a function of the Higgs mass and 
is shown in Figure~\ref{gg_xsec_limit}.  
Limits on associated Higgs production
from CDF are in progress; limits from D$\O$ and OPAL are discussed in
Refs.~\cite{D0 Bosonic Higgs,OPAL Bosonic Higgs}.

%While there is no expectation of excluding
%the Higgs using this mechanism it illustrates the sensitivity of the
%measurement to new heavy particles which decay to $\gamma\gamma$.    

%In these models, the dominant production mechanism is associated
%production (VH where V is a $W$ or a $Z^0$) and the 
%Higgs couplings to fermions is suppressed such that the
%one loop decay of the Higgs into two photons can dominate for masses less than
%around 90~GeV. 

%There are 287 %events in the sample 
%The sample is estimated to be 60\%$\pm$30\% from fakes, 
%mostly in the low mass region.  
%100$\pm$8~pb$^{-1}$ of data 

{\begin{figure}[htb!]
\vspace{-0.4cm}
 \begin{minipage}[t]{0.48\linewidth}
   \epsfxsize \linewidth
%   \hspace{0.05cm}
   \epsffile[10 128 550 675]{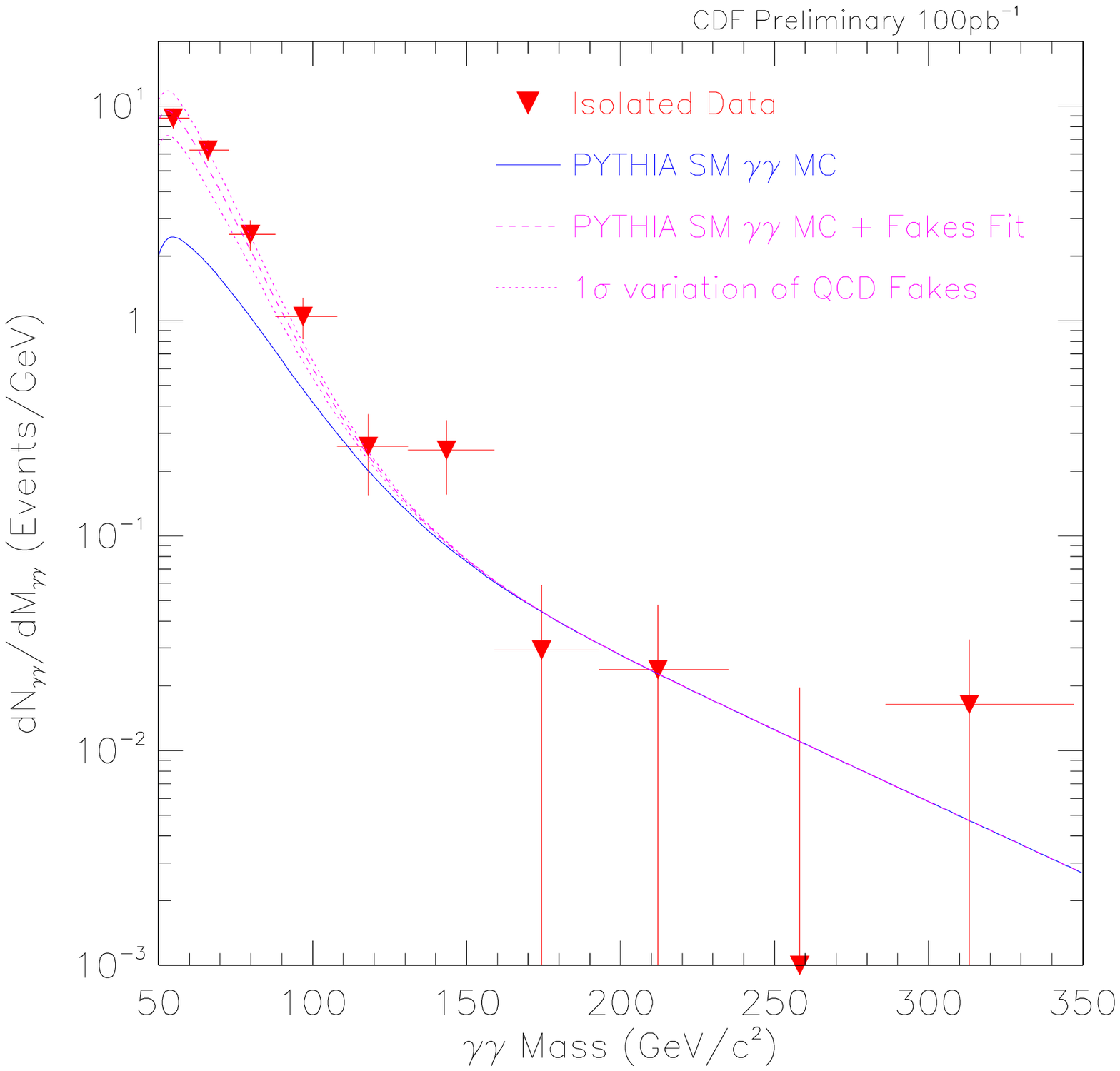}
   \vspace{2cm}
   \vspace{-0.5cm}
   \caption{Diphoton mass distribution from CDF in the high mass 
$X\rightarrow \gamma\gamma$ search. The
triangles are diphoton candidates, the solid line is the Standard Model
prediction using PYTHIA.  The dashed line is the sum of the PYTHIA diphotons
and
fakes. The dotted lines represent the $1\sigma$ uncertainties on the fakes
estimate.}
   \label{gg_xsec}
 \end{minipage}
 \hfil
 \begin{minipage}[t]{0.480\linewidth}
   \epsfxsize \linewidth
%   \hspace{2cm}
   \epsffile[10 128 550 675]{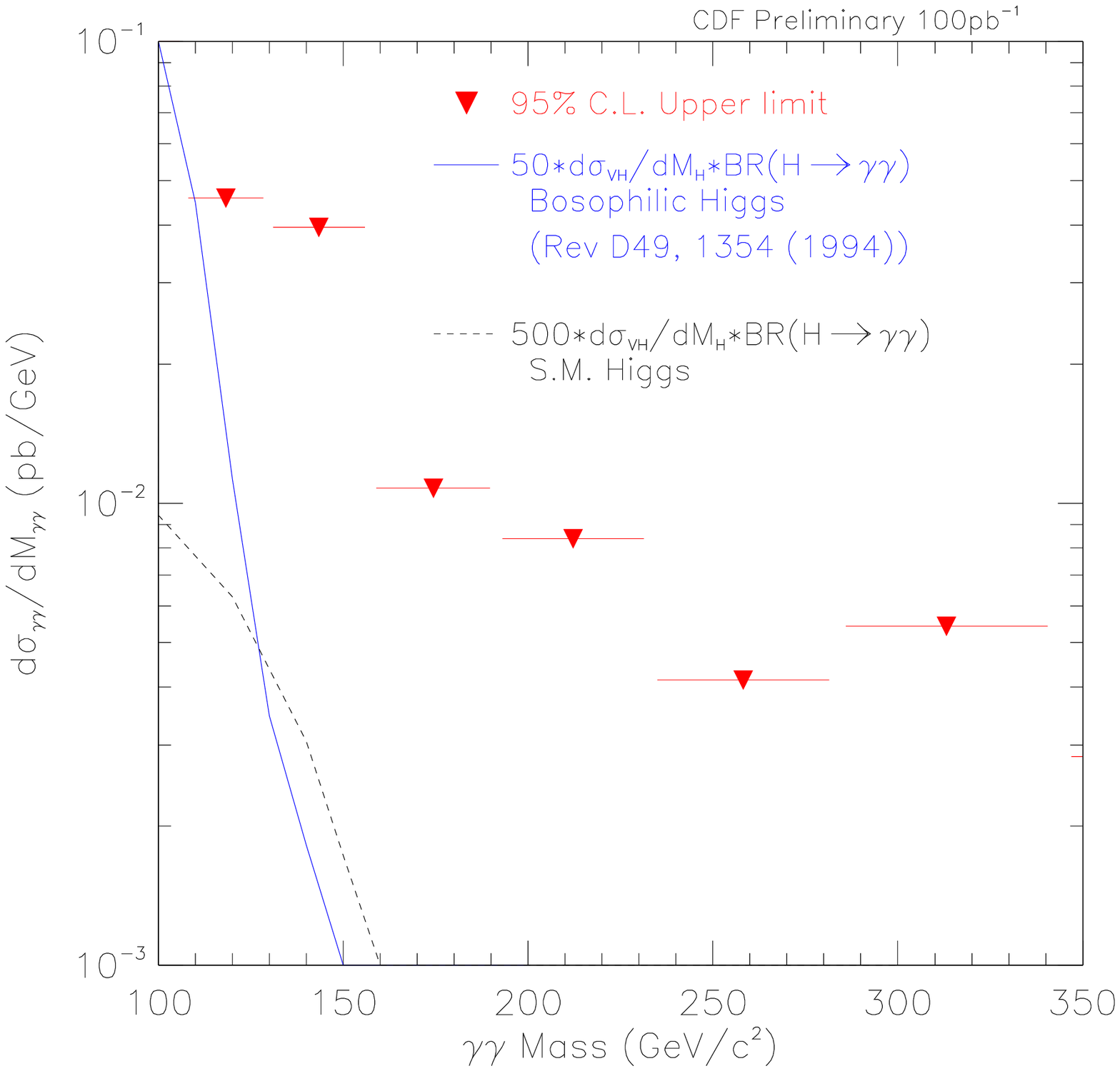}
   \vspace{2cm}
   \vspace{-0.5cm}
   \caption{Cross section limit for high mass $X\rightarrow \gamma\gamma$
production.}
% The Higgs cross sections are scaled by 20\% of the mass to
%put them into the $d\sigma/dM$ basis.}
   \label{gg_xsec_limit}
 \end{minipage}
\vspace{-1.5cm}
\end{figure}

% ******************************************************************************

\section{$\gamma\gamma + X$}

In the last couple of years, many models of new physics predict events with two 
high-P$_{\rm T}$ isolated photons produced in association with other particles
in the final state. CDF has recently submitted a paper which describes a set of
searches using 
a sample of two isolated photons with 
${\rm E_T^{\gamma}>12}$~GeV and $|\eta|<1.0$~\cite{GG PRL}. Each event 
is searched for 
anomalous production of $\gamma\gamma + X$ where $X$ is \mett\ 
(from neutrinos or
other particles such as Supersymmetric LSP's), jets, electrons, muons and taus, 
$b$-tagged jets, or extra photons.

% ******************************************************************************

\subsection{$\gamma\gamma + \mett$}

The $\gamma\gamma + \mett$ search is
optimized to look for new heavy particles produced in
Gauge-Mediated Supersymmetry with a light gravitino ($\gravitino$).
In many of these
models  the sparticles decay 
into the lightest neutralino,
$\NONE$, which in turn decays via $\NONE \rightarrow
\gamma\gravitino$ producing two photons and \mett\ in every event.
CDF has searched an integrated luminosity of 85$\pm$7~pb$^{-1}$ for
$\gamma\gamma + \mett$ events using two different
photon ${\rm E_T}$ thresholds: 
${\rm E_T^{\gamma}}>12$~GeV and ${\rm E_T^{\gamma}}>25$~GeV. 
The results are shown in Figure~\ref{CDF GGMET}.
With a threshold of E$_{\rm T}^{\gamma} > 12$~GeV, \mett$>35$~GeV, 
\NMETEXPLOW\ events are expected with \NMETLOW\  event observed.  
For a threshold of 
E$_{\rm T}^{\gamma} > 25$~GeV and  $\mett>25$~GeV, 
\NMETEXPHIGH\ events are expected with \NMETHIGH\ events observed. 
The observations are 
consistent with background expectations with one
possible exception on the tail of the distribution of 
Figure~\ref{CDF GGMET}. 
This event is the
$\eeggmett$ candidate event (see Figure~\ref{EEGGMET LEGO})
and will be discussed later.

%The D$\O$ data collected 
%represents an integrated luminosity of 106$\pm$6~pb$^{-1}$. 
%The missing transverse energy distribution is
%shown in Figure~\ref{D0 ggmet} along with the prediction from two points in
%parameter space.
%There are two events in the data with $\mett>25$~GeV 
%with an expectation of 2.3$\pm$0.9 events.

%The D$\O$ limit, interpreted in a framework of supersymmetric models with a
%light gravitino and with the assumption of gaugino mass unification at the GUT
%scale, yields a 95\% C.L. lower mass limits of 150~GeV/$c^2$ for the lightest
%chargino and 75~GeV/$c^2$ for the lightest neutralino.  Limits on
%$\tilde{\chi}^{\pm}_{1}
% \tilde{\chi}^{\pm}_{1}$ and 
%$\tilde{\chi}^{\pm}_{1} 
% \tilde{\chi}^{0}_{2}$
%production as a function of m$_{\tilde{\chi}^{\pm}_{1}}$ are shown in 
%Figure~\ref{D0 ggmet limit}.

Limits have been set on a number of  Supersymmetric models with a
light gravitino by other collaborations~\cite{D0 GG Limits, LEP GG Limits}.
The CDF limit, interpreted in a Gauge-Mediated scenario in the MSSM 
with a light gravitino and using the full one-loop normalization group 
corrections (See Ref.~\cite{Gravitino Reference}e)
is shown in
Figure~\ref{CDF ggmet limit}.  The lowest value of the
lightest chargino mass, M$_{C_1}$, that is excluded at
95\% C.L. 
is M$_{C_1} <120$~GeV. Similarly, M$_{N_1} <65$~GeV is excluded at the 
95\% C.L. 

\twofig{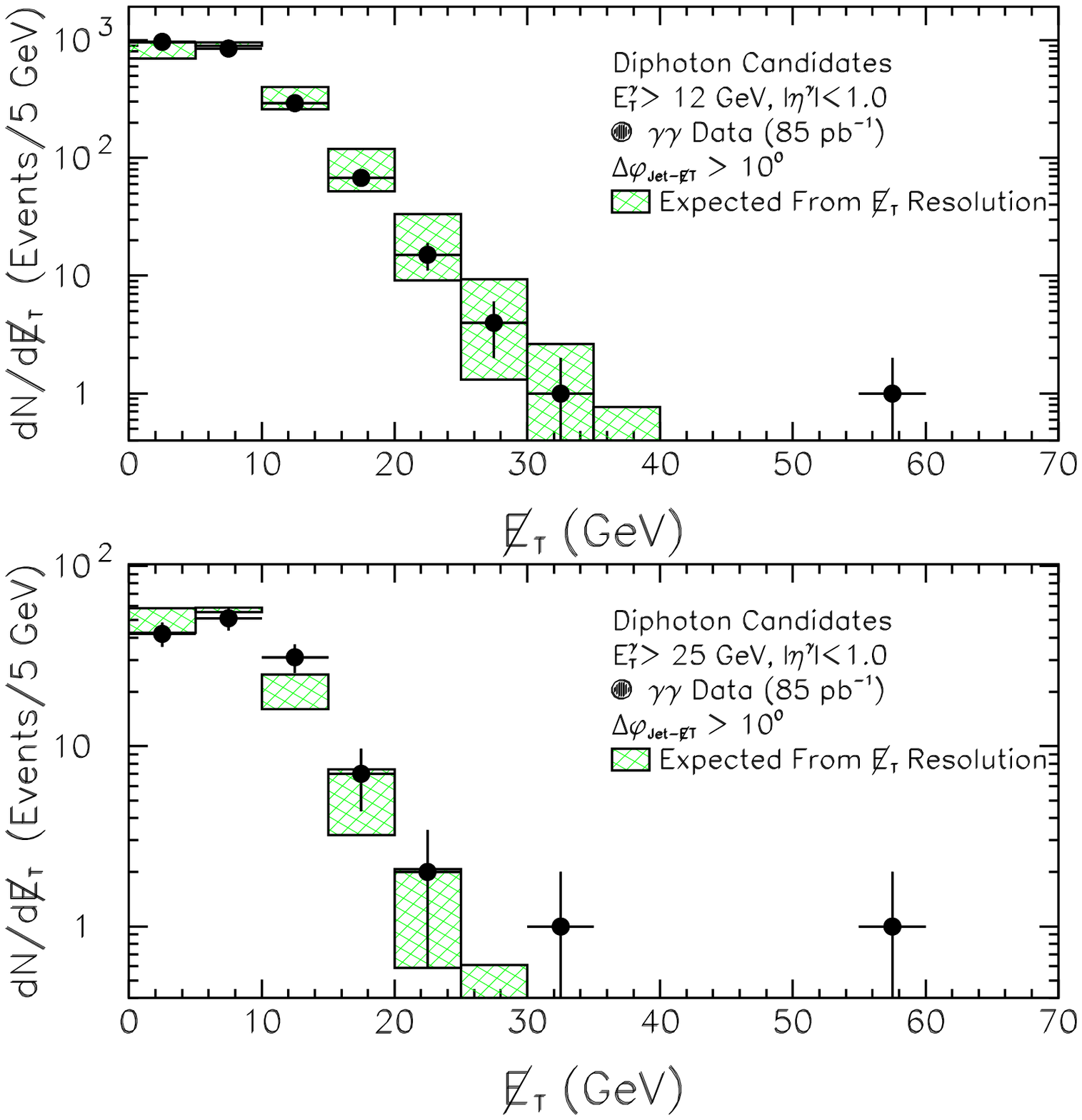}{The \mett\ spectrum for
diphoton events with \mbox{\Etggl} and \mbox{\Etggh} 
in the data from the CDF detector. 
The boxes indicate
the range of the values of the
\mett\ distribution predicted from detector resolution.
The one event on
the tail is the $\eeggmett$ candidate
event.}{CDF GGMET}{0.0cm}
{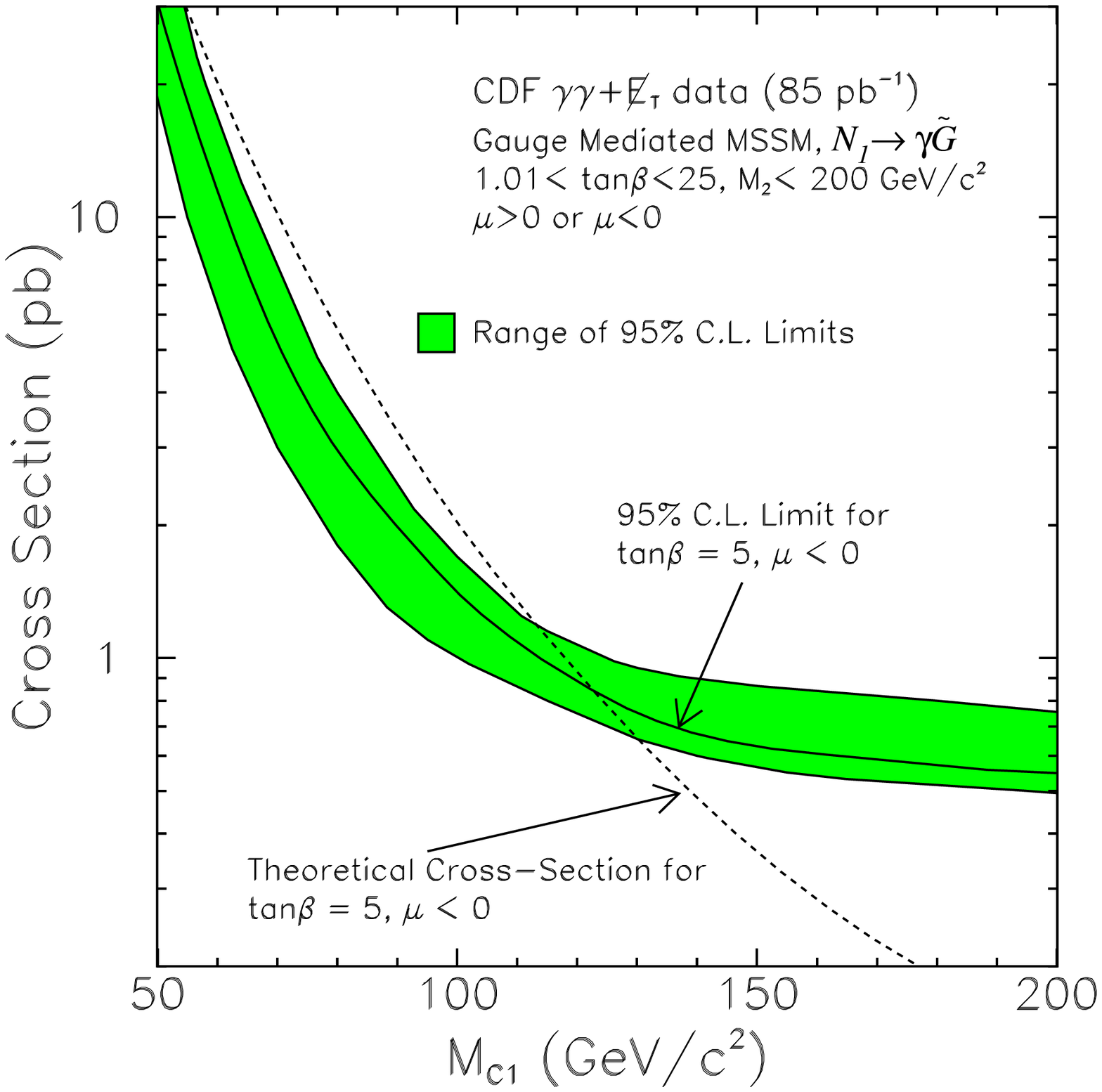}
{The cross-section upper limits versus the mass of the $C_1$
for a Gauge-Mediated model  with a 
light gravitino in the MSSM and taking into account the one-loop renormalization
group effects~\protect\cite{Gravitino Reference}e. 
The $\CONE$ is excluded for all masses with \mbox{M$_{C_1}<120$~GeV}.}
{CDF ggmet limit}{0.0cm}

%The
%shaded region shows the range of cross section limits as the parameters are
%varied within the ranges
%\mbox{$1<\tanbeta <25$}, 
%\mbox{$M_2<200~$GeV}, and \mbox{$\mu>0$} or \mbox{$\mu<0$.}
%The lines show the experimental limit and the theoretically predicted cross
%section
%for the lowest value of M$_{C_1}$ that is excluded
%(\mbox{M$_{C_1}<120$~GeV} at 95\% C.L., for $\tanbeta=5$, 
%\mbox{$\mu<0$}).}

%*****************************************************************************

\subsection{$\gamma\gamma$ + Jets}

The $\gamma\gamma$ data has also been searched for
additional jet production. 
Jets which have $\Et>$10~GeV and have $|\eta|<2.0$ are counted and compared to
expectations by using the 
low N$_{\rm Jet}$ results to estimate the production for large
N$_{\rm Jet}$. In the data with two photons with 
E$_{\rm T}^{\gamma} > 12$~GeV, 2 events are observed with 
$\ge$ 4 jets with \NIVJETEXPLOW\ expected. Similarly, when both
photons are required 
to have E$_{\rm T}^{\gamma} > 25$~GeV, a total of $\NIIIJETEXPHIGH$ are expected
with $\ge$ 3 jets with $\NIIIJETHIGH$  events observed. These results are shown
in Figure~\ref{CDF NJet} and are consistent with background expectations.
%As previously mentioned, the D$\O$ and OPAL
%collaboration have looked for  associated bosonic Higgs production with
%$VH \rightarrow \gamma\gamma + jj$~\cite{D0 Bosonic Higgs, OPAL Bosonic Higgs} 
%and CDF results in this mode are
%expected shortly. 

%This search uses 101$\pm$6~pb$^{-1}$ of data. In this study, events with
%{${\rm E_T^{\gamma_1}>20}$~GeV and $|\eta_1|<2.0$,} 
%{${\rm E_T^{\gamma_1}>15}$~GeV and $|\eta_2|<2.25$} are searched 
%for the hadronic
%decays of the $W$ and $Z^0$ bosons. The jets are required to have 
%{${\rm E_T^{Jet_1}>20}$~GeV and $|\eta_1|<2.0$,}
%{${\rm E_T^{Jet_2}>15}$~GeV and $|\eta_2|<2.25$}.  
%Figure~\ref{D0 ggjj}
%shows the diphoton invariant mass for events which have 
%40~GeV$\leq {\rm M_{jj}} \leq 150$~GeV. 
%A background estimate, dominated by fake photons, is 
%10.5$\pm$4.0 events expected with 7 observed. For a 
%M$_{\gamma\gamma}>60$~GeV 3.5$\pm$1.3, while zero are seen.  
%Figure~\ref{D0 ggjj Limit} shows the cross
%section theory predictions as well as the 90\%
%and 95\% C.L. cross section upper limits.

%The jet and photon pairs are both required to have a system $\Et>$10~GeV. 
%Since there is no excess of events, limits are set. 

\twofig{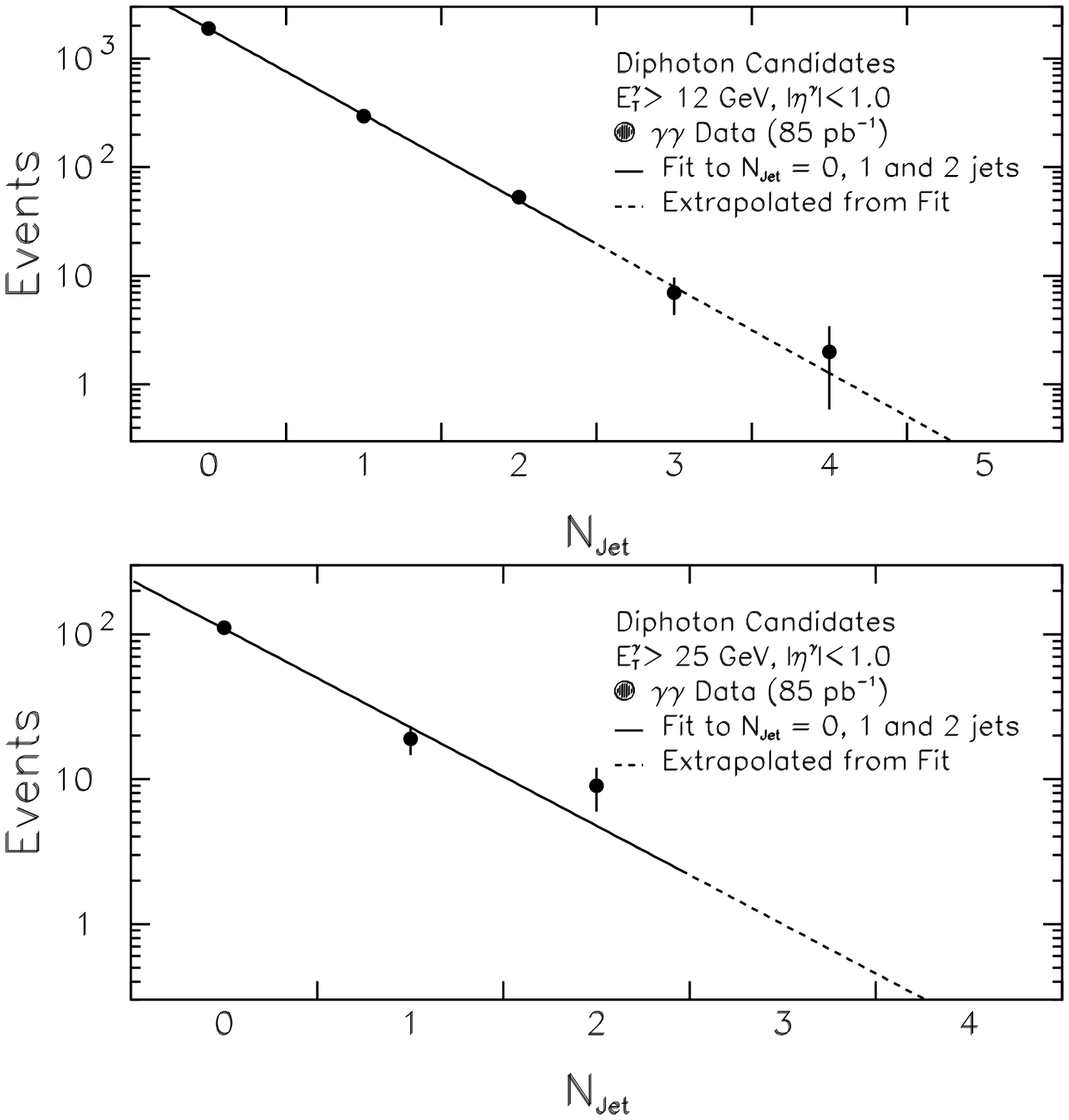}
{The number of jets, N$_{\rm Jet}$, produced in association with
diphoton pairs with \Etggl\ and \Etggh.
The solid 
line is an exponential fit to the data, the dashed line represents the
extrapolation for large values of N$_{\rm Jet}.$}{CDF NJet}{0.0cm}
{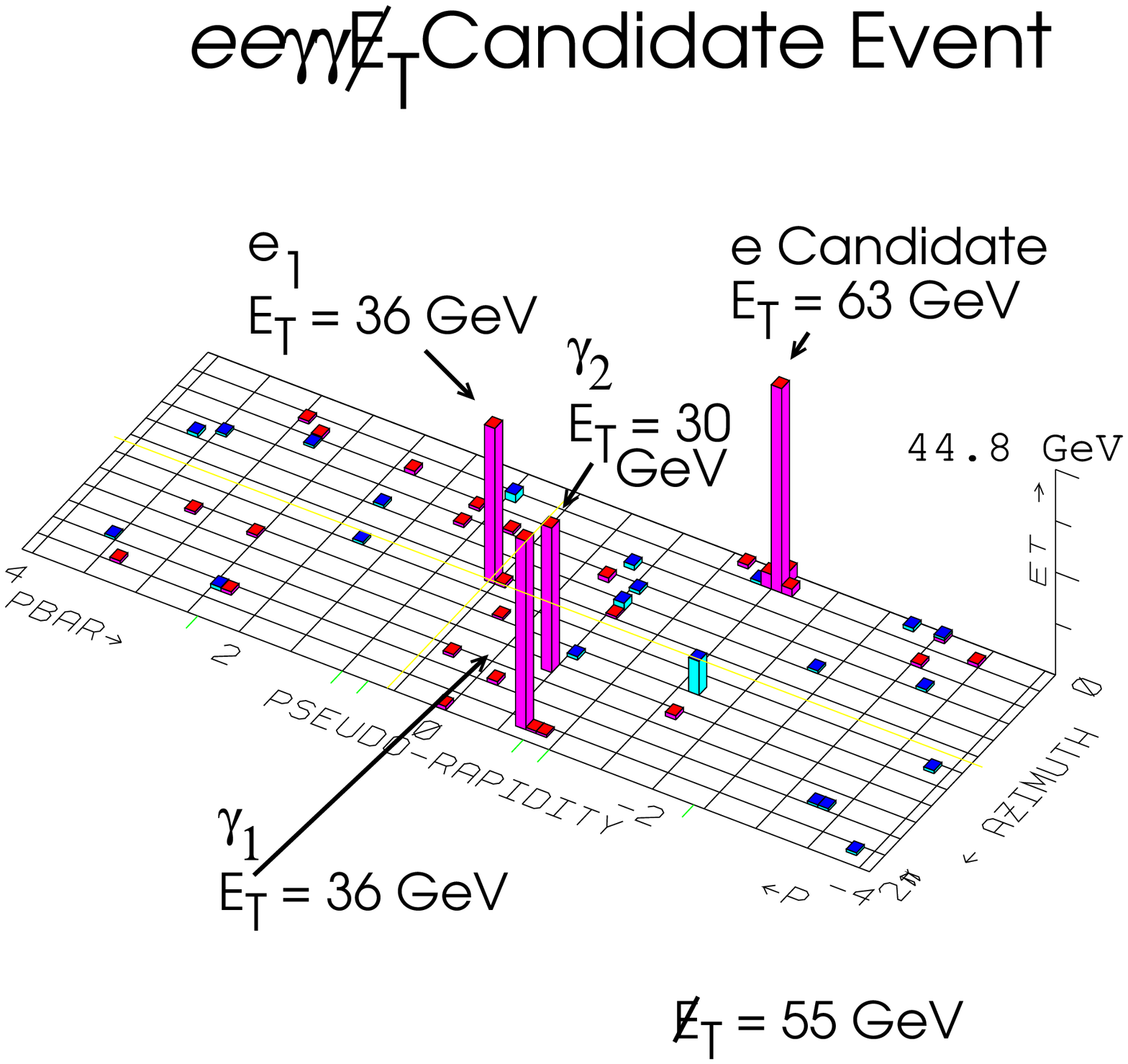
}{An event display for the CDF
$\eeggmett$ candidate event.}{EEGGMET LEGO}{0.0cm}

% ***************************************************************

\subsection{Other $\gamma\gamma + X$ Results}

The $\gamma\gamma$ data is also searched for the presence of
associated lepton, \mbox{$b$-quark} and/or additional
photon production. 
The results are presented in Table~\ref{found}.
A total of \NBTAG\ events  with $b$-tags are found, 
consistent with background expectations. No events with a third photon are
found. There are 4 events with a central
lepton: one event is consistent with a double-radiative $\Z$ decay
($m_{\mu\mu\gamma\gamma}=\MUMUGGMASS$~GeV/c$^2$), one is consistent with a
radiative $\Z$ decay with a lost track ($m_{ee\gamma}=\EGGMASS $~GeV/c$^2$),
and one has a $\tau$ candidate, for which there is a fake background 
expectation of of
\NTAUEXPLOW\ events.  Other than the fourth event, the $\eeggmett$
candidate event,
there is 
agreement between the observations and the predictions. 

\begin{table}[t!]
\vspace{1.5cm}
\centering
\begin{tabular}{lcc}
\hline
\multicolumn{3}{c}{\Etggl\ Threshold} \\
\hline
Signature (Object) & Obs. & Expected  \\
\hline
\mettgmh, $|\Delta\phi_{\mettsm-{\rm jet}}|>10^\circ$
                           & \NMETLOW      & \NMETEXPLOW           \\
N$_{\rm jet}\ge 4$, ${\rm E}_{\rm T}^{\rm jet}>10$~GeV,
$|\eta^{\rm jet}|<2.0$     & \NIVJETLOW    & \NIVJETEXPLOW    \\
Central $e$ or $\mu$, ${\rm E}_{\rm T}^{e~{\rm or}~\mu}>25$~GeV
    & \NCENTEORMU   & \NCENTEORMUEXP        \\
Central $\tau$, ${\rm E}_{\rm T}^{\tau}>25$~GeV
          & \NCENTTAU     & \NTAUEXPLOW \\
$b$-tag, ${\rm E}_{\rm T}^{b}>25$~GeV
                  & \NBTAG        & \NBEXPLOW \\
Central $\gamma$, ${\rm E}_{\rm T}^{\gamma_3}>25$~GeV
                        & \NADDGAMMA    & \NADDGAMMAEXP         \\
%\hline
%\Etggh, \mettgl         & \NMETHIGH     & \NMETEXPHIGH          & -- \\
%\Etggh, N$_{\rm jet} \ge 3$
%                        & \NIIIJETHIGH  & \NIIIJETEXPHIGH       & \cite{top} \\
\hline\hline
\multicolumn{3}{c}{\Etggh\ Threshold} \\
\hline
Object & Obs. & Exp.  \\
\hline
\mettgl,           $|\Delta\phi_{\mettsm-{\rm jet}}|>10^\circ$
         & \NMETHIGH     & \NMETEXPHIGH          \\
N$_{\rm Jet} \ge 3$, ${\rm E}_{\rm T}^{\rm Jet} >10$~GeV,
$|\eta^{\rm Jet}| < 2.0$ & \NIIIJETHIGH  & \NIIIJETEXPHIGH \\ 

Central $e$ or $\mu$, ${\rm E}_{\rm T}^{e~{\rm or}~\mu} >25$~GeV
    & 1   & 0.1 $\pm$ 0.1 \\
Central $\tau$, ${\rm E}_{\rm T}^{\tau} >25$~GeV
          & 0     & 0.03 $\pm$ 0.03         \\
$b$-tag, ${\rm E}_{\rm T}^{b}>25$~GeV
                  & 0        & 0.1 $\pm$ 0.1            \\
Central $\gamma$, ${\rm E}_{\rm T}^{\gamma_3}>25$~GeV
                        & 0    & 0.01 $\pm$ 0.01      \\ 
\hline\hline
\end{tabular}
\caption{Number of observed and expected  $\gamma\gamma$ events with additional
objects in 85 pb$^{-1}$ from CDF.}
\label{found}
\vspace{-2cm}
\end{table}

% ***************************************************************
\section{The CDF $\eeggmett$ Candidate Event}

\begin{sloppypar}The CDF $\eeggmett$ candidate event~\cite{GG PRL} 
(see Figure~\ref{EEGGMET LEGO}) is an unusual event.
It has two photons which are very energetic (E$_{\rm T}$ = 32 and 36~GeV 
respectively), large $\mett$ (\mbox{$\mett= 55\pm 7$~GeV}), a 
high-E$_{\rm T}$ central electron  (36~GeV) and an 
electromagnetic cluster in the plug calorimeter (63~GeV) which easily 
passes the standard electron selection criteria used for $\Z$ 
identification. 
The total \ptt\ of the 4-cluster system is 
\mbox{$\EEGGPT$~GeV/c,} opposite to the \mett\
and in good agreement with the measured magnitude, implying  the imbalance is
intrinsic to the 4-cluster system. 
The invariant mass of the electron and the
electromagnetic cluster  in the plug calorimeter is \mbox{$\EEMASS$~GeV/c$^2$},
far from the $\Z$ mass. 
\end{sloppypar}

 \dtfloat{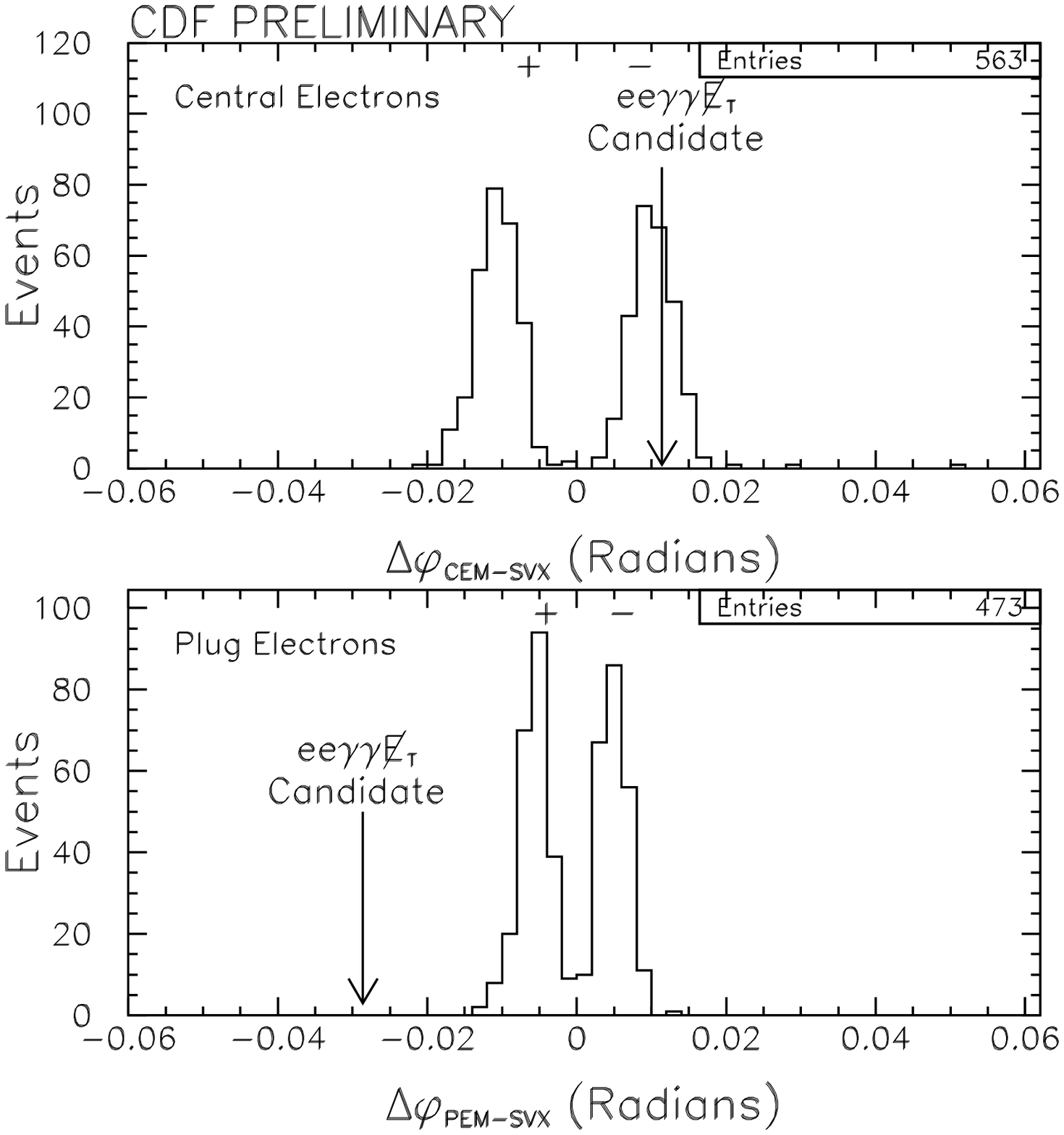}
{The $\Delta\phi$ between the measured electron position from the
calorimeter and the $\phi$ from the SVX tracker for electrons from a $\zoee$
control sample. The
two peaks  correspond to the bending of positively and negatively charged
electrons in the magnetic field.}
{Delta Phi Plot}{0.440}

Although the electromagnetic cluster  in the plug calorimeter passes all of the
standard electron selection criteria,  
there is reason to believe it is not an electron. CDF uses three
tracking chambers to determine the charged particle trajectory. 
The trajectory only passes through the inner 
layers of the Central Tracking Chamber, so the
efficiency to see a track is $\approx 0$. In addition, 
there is too much activity from the underlying event to see individual
hits which might indicate a track. 
The results from the vertex finding chamber 
(VTX) are completely consistent with the single electron hypothesis. However,
these results are not
as precise as the silicon vertex detector (SVX) which shows 
no clear track pointing directly at the cluster. 
There is, however,  a track
26~mrad away  in $\phi$.  While this electron may have passed through a bad
region of the detector, there are no other such events in a 
control sample of electrons (see Figure~\ref{Delta Phi Plot}) from 
$\zoee$ events. The probability of an electron to have a 
$\phi$ mismatch this large is estimated to be 
less than 0.3\% at 95\%~C.L.

\begin{sloppypar}
The interpretation of the cluster as coming
from a photon, the 1-prong hadronic decay of a $\tau$, or a jet, while
possible, are unlikely in that this would be an unusual example of any
of them~\cite{Future PRD}. 
The photon interpretation  is  unlikely as it requires the 
tracks in the SVX and VTX to be completely unrelated. 
The probabilities of this have
been estimated to be at the few percent level. 
Another possibility is that this is
the 1-prong hadronic decay of a $\tau$,  e.g, 
$\tau~\rightarrow~\pi^+\pi^0\nu_{\tau}$. However, more than 95\% of
$\tau$'s leave a greater percentage of their 
energy in hadronic calorimeter than is observed.
The probability that a generic jet would fake the 
electron signature in the plug calorimeter
is estimated to be \mbox{\scinotn{2}{-3} / jet. }
The conclusion is that 
there is not
enough  information to establish the origin of
the cluster.
\end{sloppypar}

\clearpage
% ***************************************************************

\subsection{How many $\eeggmett$ events expected?} 

The Standard Model  rates for producing a signature of  two
photons, two electromagnetic clusters (one central) passing the electron
requirements
and $\mett$, all with \mbox{$\Et>25$~GeV},
and \mbox{M$_{ee}>110$~GeV/c$^2$} (above the
$\Z$ boson)~\cite{Future PRD} has been estimated for 
 $WW\gamma\gamma$ and $t{\bar t}$, as well as sources which include additional
cosmic ray interactions, jets which fake electrons and/or photons,
and overlapping events.

\begin{sloppypar}
The primary estimate allows sources from both 
real electrons and fakes to contribute. The number of events is dominated
by $WW\gamma\gamma$ production 
which should produce \scinotn{(8\pm 8)}{-7} events. The next highest
producer is $t{\bar t}$ and yields \scinotn{(3\pm 3)}{-7} events. 
Fakes contribute  \scinotn{(3\pm 3)}{-7} events. Two 
overlapping events (including cosmic ray contributions) contribute 
\scinotn{(8\pm 8)}{-9} events.  
Summing all sources
yields \scinotn{(1\pm 1)}{-6} events.  Since the plug cluster may not be due to
an electron, a separate estimate is made by  
including only sources where the plug cluster is due to a jet faking
an electron.
The rate is reduced because the dominant
sources are no longer allowed to contribute and   the dominant source
becomes 
$e\gamma\gamma\mett + j$ where the jet fakes the
plug electron. In this case, the total rate is estimated to be 
\EEFAKEGGMETTOTRATE\ events. 
\end{sloppypar}
% ***************************************************************
\section{Interpreting the Event}

There has been a great deal of speculation about the $\eeggmett$ candidate
event. While {\it a priori} it's unlikely to be $WW\gamma\gamma$ production, it
could be an example of `anomalous' $WW\gamma\gamma$ 
production. This hypothesis allows for a 
quantitative estimate of anomalous $WW\gamma\gamma$ by looking for the 
hadronic decays of $WW\gamma\gamma$. 
%Using the equation:
%\begin{eqnarray*}
%{N_{\gamma\gamma jjj}^{\rm Expected} & \approx &
%N_{\gamma\gamma \ell_i\ell_j +\mettsm}^{\rm Observed} \times
%(\frac{{\rm Rate}~(WW\gamma\gamma \rightarrow \gamma\gamma jjj)}
%   {{\rm Rate}~(WW\gamma\gamma \rightarrow \gamma\gamma \ell_i\ell_j +\mett)})
%}
%\end{eqnarray*}
Given 1 $\ell\ell\gamma\gamma\mett$ candidate event and using the ratio of
acceptances and branching ratios, CDF predicts a total of 
30 times as many $\gamma\gamma jjj$ events to be seen in the data with two
photons with E$_{\rm T}>25$~GeV. From Figure~\ref{CDF NJet} no such
$\gamma\gamma jjj$ events are seen in the data. 
Anomalous $WW\gamma\gamma$ is excluded at the 95\% C.L. as the source of the
$\eeggmett$ candidate event.

%\subsection{SUSY or Other New Physics?}

Most new models which predict $\eeggmett$ events 
also predict $\gamma\gamma + \mett$. Currently a number of 
limits have already been 
for these models and there is no experimental 
evidence which furthers the understanding of the
event. 

%However, there are many other possibilities 
%which remain untested. Some of these 
%are presented in the next section.

%These types of searches are in progress and . One other
%possibility is 
%\begin{equation}
%$p{\bar p} \rightarrow C_iN_2 \rightarrow (b\stop)(\gamma\NONE) \rightarrow$}\\
%{$b(c\NONE)\gamma\NONE \rightarrow \gamma + b + \mett + X$}
%\end{equation}
%This type of analysis is 
%done by the CDF collaboration and a search is conducted
%in $\gamma + b + X$. This is discussed in the next section.

%*****************************************************************************

\clearpage
\section{Searches in $\gamma b + X$}

\subsection{SUSY Models: light $\tilde{t}$}

Motivated in part 
by the $\eeggmett$ candidate event, Ambrosanio \etal\ have 
predicted a neutralino
LSP model, $\NTGNO$~\cite{Higgsino LSP} with a light $\tilde{t}$. 
In this case, a chargino would decay via $C \rightarrow b \tilde{t} \rightarrow
b (c\NONE)$ and $C\NTWO$ production would have the final state signature $\
\gamma bc \mett$. At the Tevatron, the dominant production
mechanism is from squark and gluino production which decay to 
$C\NTWO$. CDF has studied $\gamma + b$
production in 85$\pm$7~pb$^{-1}$ of data with isolated photons with ${\rm
E_T}>25$~GeV and $|\eta|<1.0$, and a $b$-tagged jet 
with ${\rm E_T}>30$~GeV, and 
$|\eta|<2.0$. A total of 1175
events are observed with 1000$\pm$200 events expected, dominated by 
multijets with fake photons. 
The \mett\ distribution is shown in Figure~\ref{gamma b met data} with
two events passing 
a cut of $\mett>40$~GeV.
As a second search, the N$_{\rm Jet}$ 
distribution in the $\gamma + b +\mett$ events is studied. 
The N$_{\rm Jet}$ distribution
for events with \mett$>$20~GeV is shown in Figure~\ref{gamma b met data}. 
Since there is no excess, 
limits are set and shown in Figure~\ref{gamma b mett Limit}.

%, a total of 2.3$\pm$1.6 events are expected and 
%two events remain in the data. 

%. There are a total of 
%98 events in the data with 80$\pm 30$#
%%23 (stat) \pm 20 (syst)$ 
%events expected; 
%the N$_{\rm Jet}$ distribution

%set using the \mett\ distribution. 
%No background subtractions is done and limits are set using the two observed
%events. The limits are

%After requiring $\Delta\phi(\gamma-\mett)<2.93$

{\begin{figure}[tb]
 \vspace{2.5cm}
 \begin{minipage}[t]{0.48\linewidth}
   \epsfxsize \linewidth
%   \hspace{0.05cm}
\vspace{-1.3cm}
   \epsffile[130 415 480 590]{
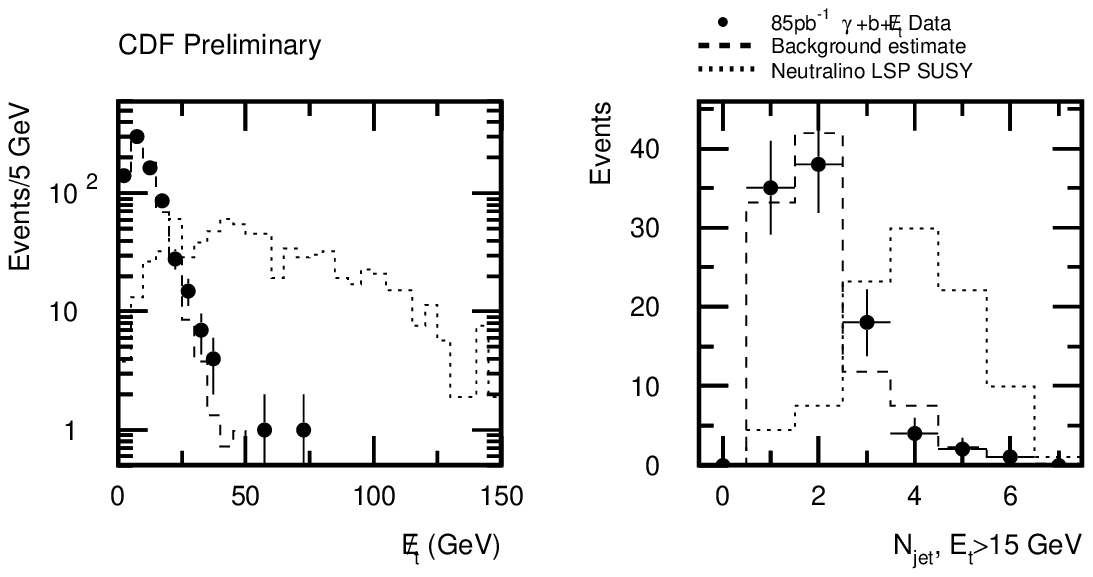}
   \vspace{-0.5cm}
   \caption{The missing ${\rm E_T}$ in $\gamma+b$ events, and the number of 
jets produced in $\gamma+b+\mett$ events from CDF. 
 The dashed 
line is the background prediction, the points are the data, and 
the dotted line
is the prediction from the light $\stop$ model. 
The N$_{\rm Jet}$ spectrum is made by requiring $\mett>20$~GeV. The SUSY model
is normalized by a factor of 100 times the expected rate in the \mett\ plot and
a factor of 10 in the N$_{\rm Jet}$ plot.}
   \label{gamma b met data}
 \end{minipage}
 \hfil
 \begin{minipage}[t]{0.480\linewidth}
   \epsfxsize \linewidth
   \vspace{-1.5cm}
   \epsffile[124 269 494 575]{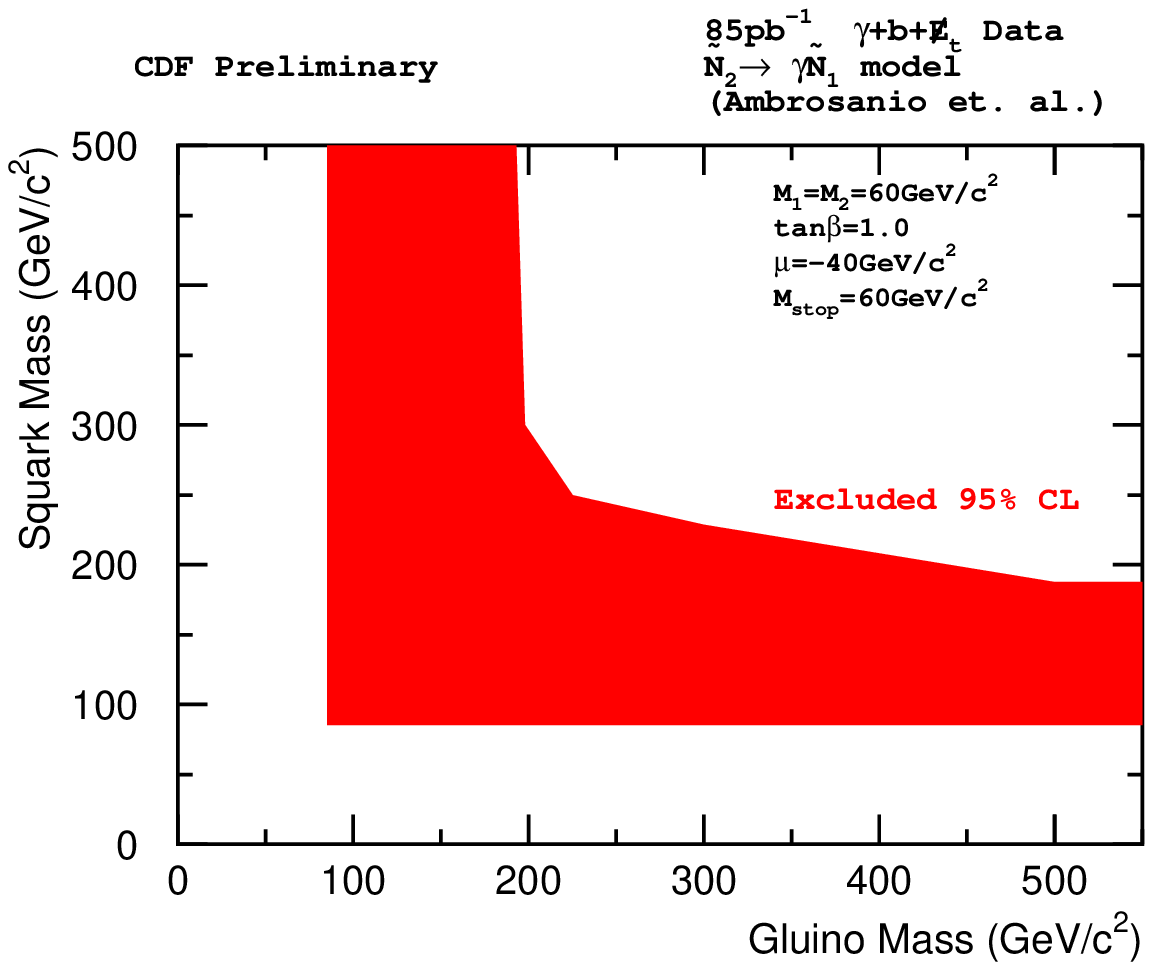}
   \vspace{-0.5cm}
   \caption{The area in the squark and gluino mass plane excluded in the
   $\NTGNO$
model of Ambrosanio \etal, using the $\gamma+b+\mett$ analysis.}
   \label{gamma b mett Limit}
 \end{minipage}
\vspace{-1.5cm}
\end{figure}

%**************************************************************************
\subsection{Technicolor: $\omega_{\rm T}$}

In new models of Technicolor~\cite{Techni Omega}, 
there exists a new particle,
$\omega_{\rm T}$, which for some region of parameter space decays
predominantly via $\omega_{\rm T} 
\rightarrow \gamma \pi_{\rm T} \rightarrow \gamma b{\bar b}$.
CDF has searched
for such a decay using the same $\gamma + b$ datasets as before. 
In addition to the
isolated photon with ${\rm E_T}>25$~GeV and $|\eta|<1.0$, two jets are required
with ${\rm E_T}>30$~GeV and \mbox{$|\eta|<2.0$,} one of which is required to be
$b$-tagged.  There are 200 events in the data sample with 
130$\pm40$ events expected, dominated by fake photon backgrounds.
Since the masses are fully reconstructed,  peaks are expected in the 
 M(b,jet) and 
\mbox{M($\gamma$,b,jet) - M(b,jet)} distributions. 
The data as well as background
expectations are shown in Figure~\ref{TECH_MASS}. There is no evidence of new
particle production. Limits are set on production assuming Br$(\omega_{\rm T}
\rightarrow \gamma b{\bar b}) = 100\%$ with the results shown in
Figure~\ref{Techni Limits}.

%131$\pm
%30(syst) \pm 29 (stat)$ events expected, dominated by fake photon backgrounds.

%\footnote{Note that
%we have assumed that $\omega_{\rm T} \rightarrow 3\pi, 2\pi and or Z\pi, f{\bar
%f}$ are negligible here.}.

\twofig{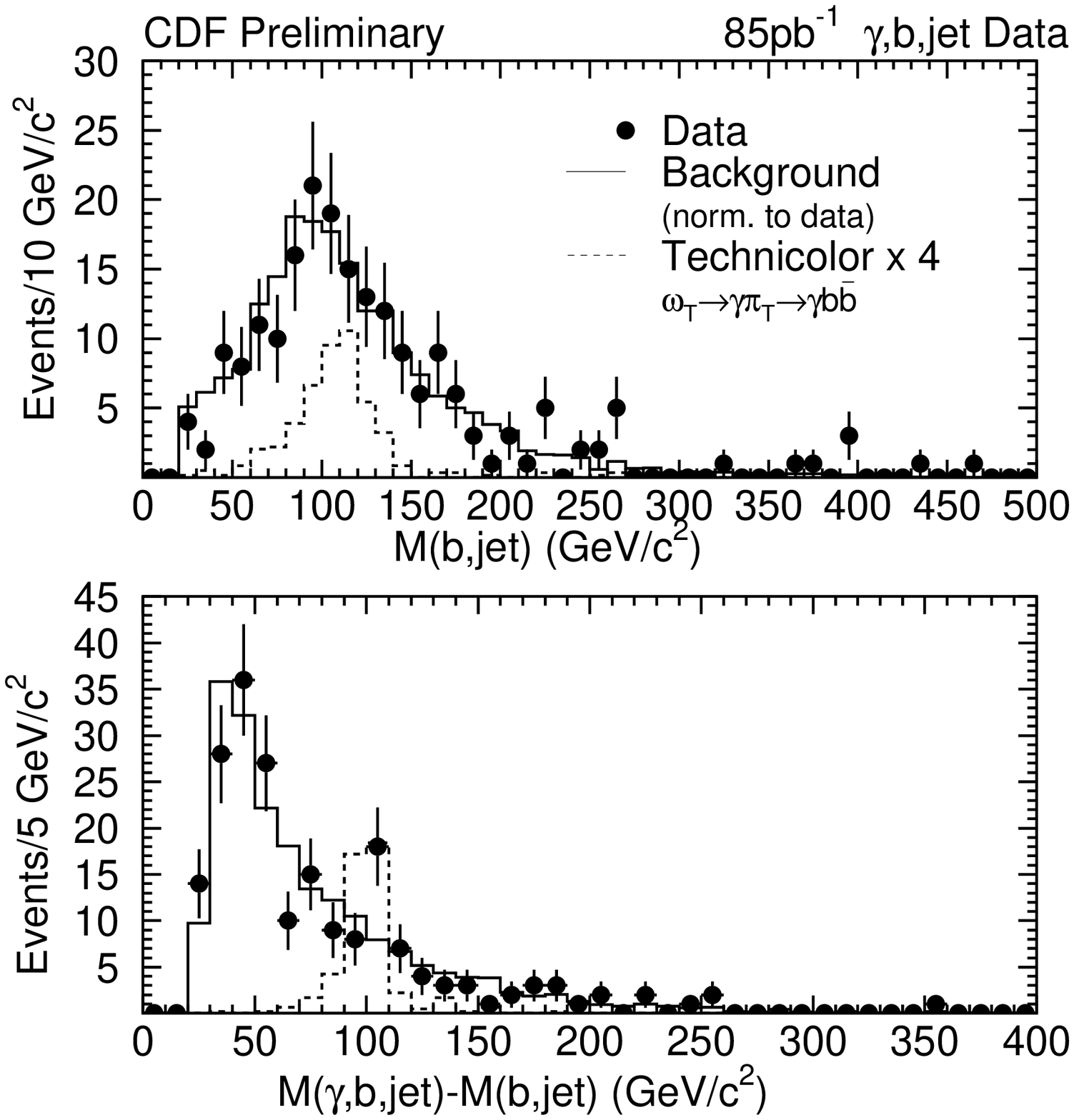}
{The M(b,jet) (top) and 
\mbox{M($\gamma$,b,jet) - M(b,jet)} (bottom)
for events with a photon, one $b$-tagged jet and second jet. The
data are the points with error bars; the solid histogram is the background
estimate (normalized to the data); the dashed histogram is a Monte Carlo
simulation of a point in Technicolor parameter space, scaled up by a factor of
4}{TECH_MASS}{0.0cm}
{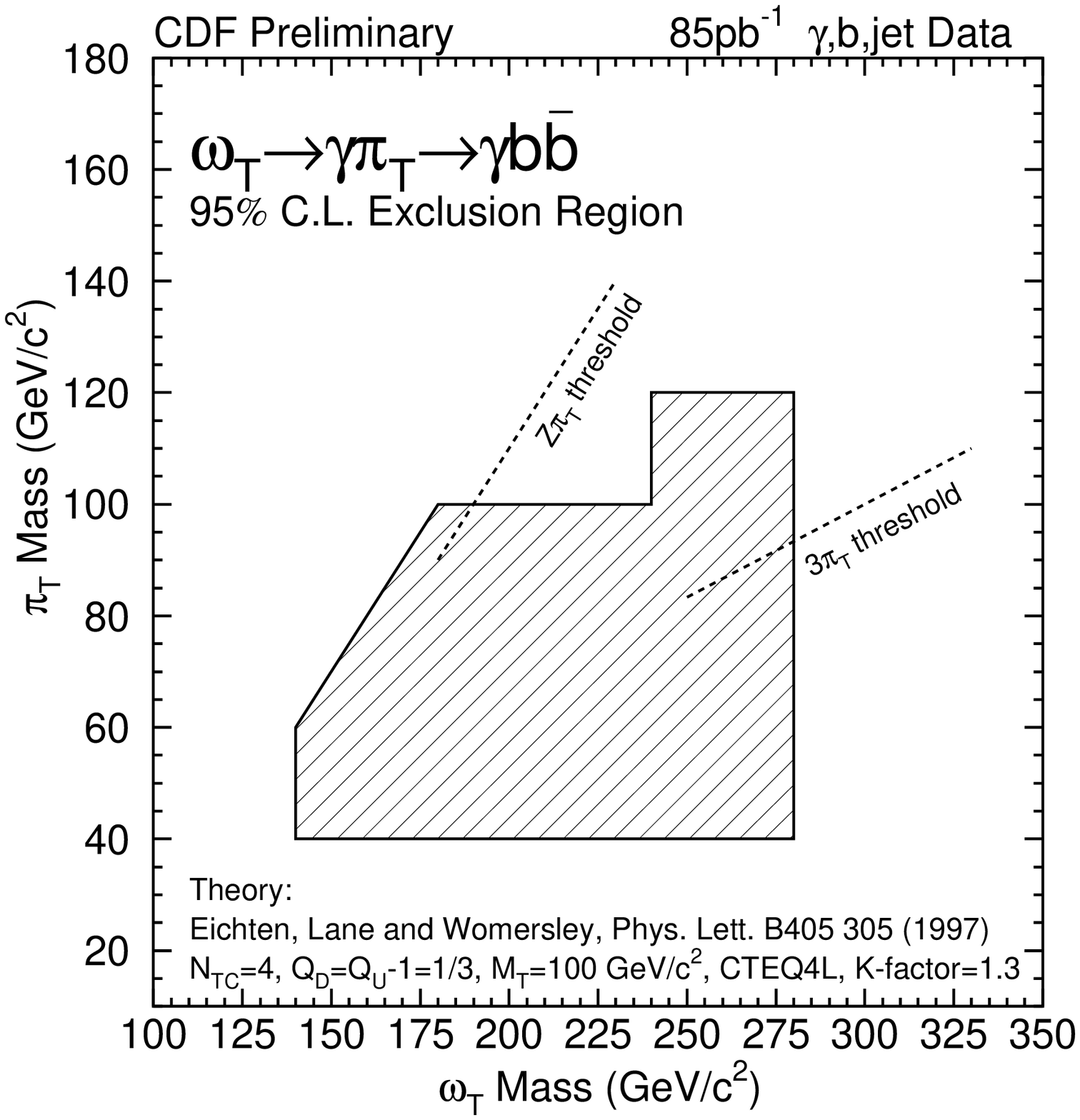}
{The 95\% C.L. excluded region in the
$\omega_{\rm T} - \pi_{\rm T}$ mass plane. The dashed lines indicate the
thresholds for decay modes that are assumed to be zero in the model.}
{Techni Limits}{0.0cm}

% ***************************************************************

\section{$WX$ and $ZX$ production}

\subsection{Technicolor: $\rho_{\rm T}$}

The $\omega_{\rm T}$ is not the only Technicolor particle recently searched for
at the Tevatron. Recent results from a 
CDF search in 109$\pm$7~pb$^{-1}$ of data look
for the production and decay of 
the $\rho_{\rm T}$. The $\rho_{\rm T}$ can come in charged and neutral states
and decay via $\rho_{\rm T}^0
\rightarrow W \pi_{\rm T}^{\pm} \rightarrow \ell\nu
b{\bar c}$ and 
$\rho_{\rm T}^{\pm}
\rightarrow W \pi_{\rm T}^0 \rightarrow \ell\nu
b{\bar b}$~\cite{Techni Rho}. The search requires a high-P$_{\rm T}$,
charged lepton ($e$ or $\mu$), missing transverse energy, and two jets, one 
of which is required to pass $b$-tagging requirements.  The $e$ or $\mu$ is
required to be isolated, have ${\rm P_T}>20$~GeV and have $|\eta|<1.0$. The
missing transverse energy is required to be $\mett > 20$~GeV. The event is
required to have two and only two jets with ${\rm E_T}> 15$~GeV and 
$|\eta|<2.0$.
A total of 42 events are observed in the data with an expectation of 32$\pm$4
events.
The dijet and $W+$dijet masses\footnote{The $W$+Dijet mass is made by
constraining the $\ell\mett$ system to have a mass consistent with that of a $W$
boson.} are shown in Figure~\ref{Technirho}. While a 
signal would show up as a bump in both distributions, there 
is no evidence for a bump in either. Limits on the production of the
$\rho_{\rm T}$ are shown in Figure~\ref{Trho Limits}.

\twofig{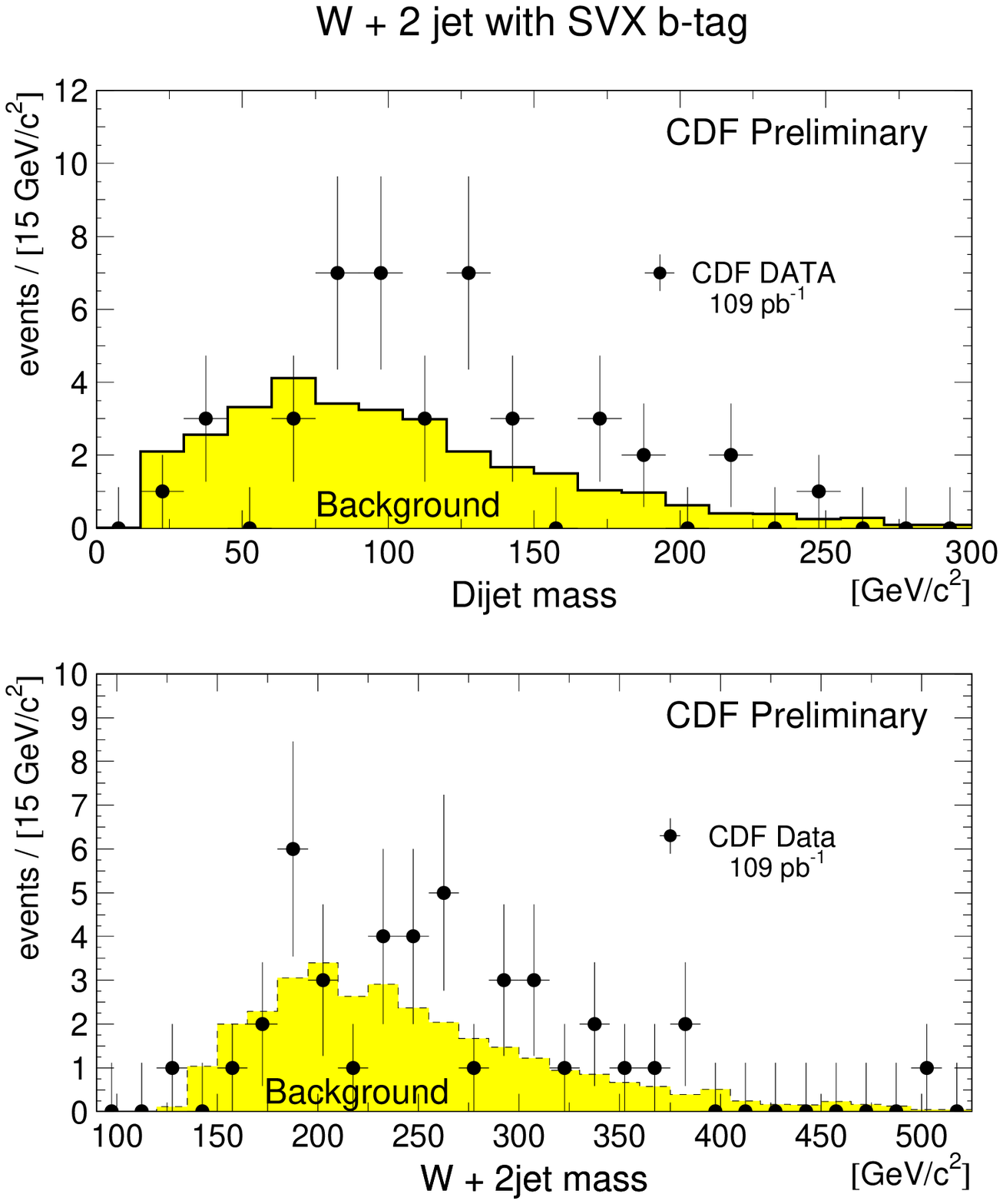}
{The invariant mass of the dijet and 
$W$+dijet mass systems for
the $W$+2 jet sample with a $b$-tag.}
{Technirho}{0.0cm}
{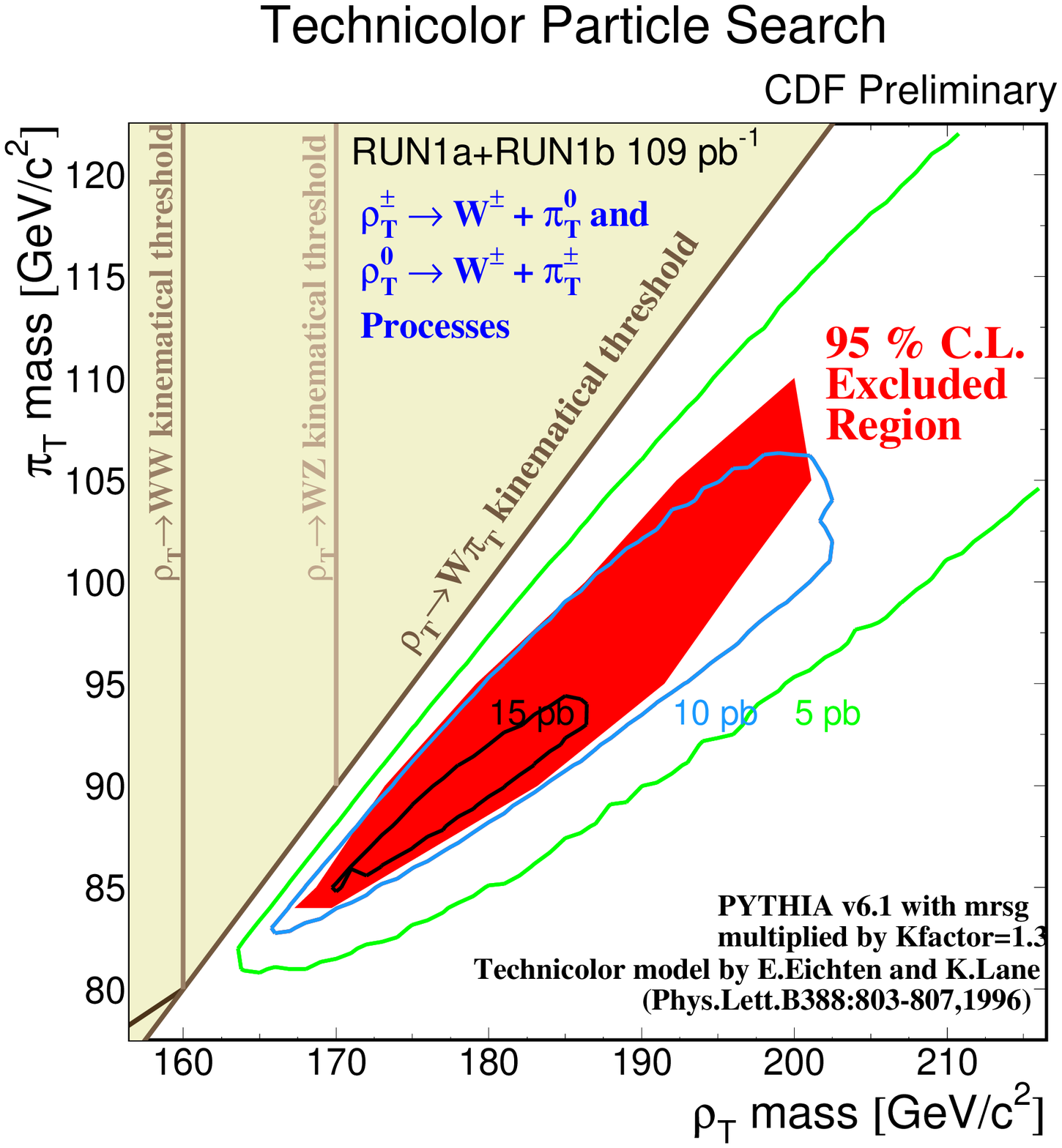}
{The 95\% C.L. excluded region in the 
M$_{\pi_{\rm T}}$ vs. M$_{\rho_{\rm T}}$ mass plane (dark 
shaded region). The lines represent the contour of constant total 
production cross section.}
{Trho Limits}{0.0cm}

%************************************************************************

\subsection{Associated Higgs $\rightarrow b{\bar b}$}

\begin{floatingfigure}{0.48\linewidth}
%\begin{figure}
\vspace*{-4cm}
\hspace*{-0.5cm}
\epsfxsize 0.48\linewidth
   \epsffile[10 128 550 675]{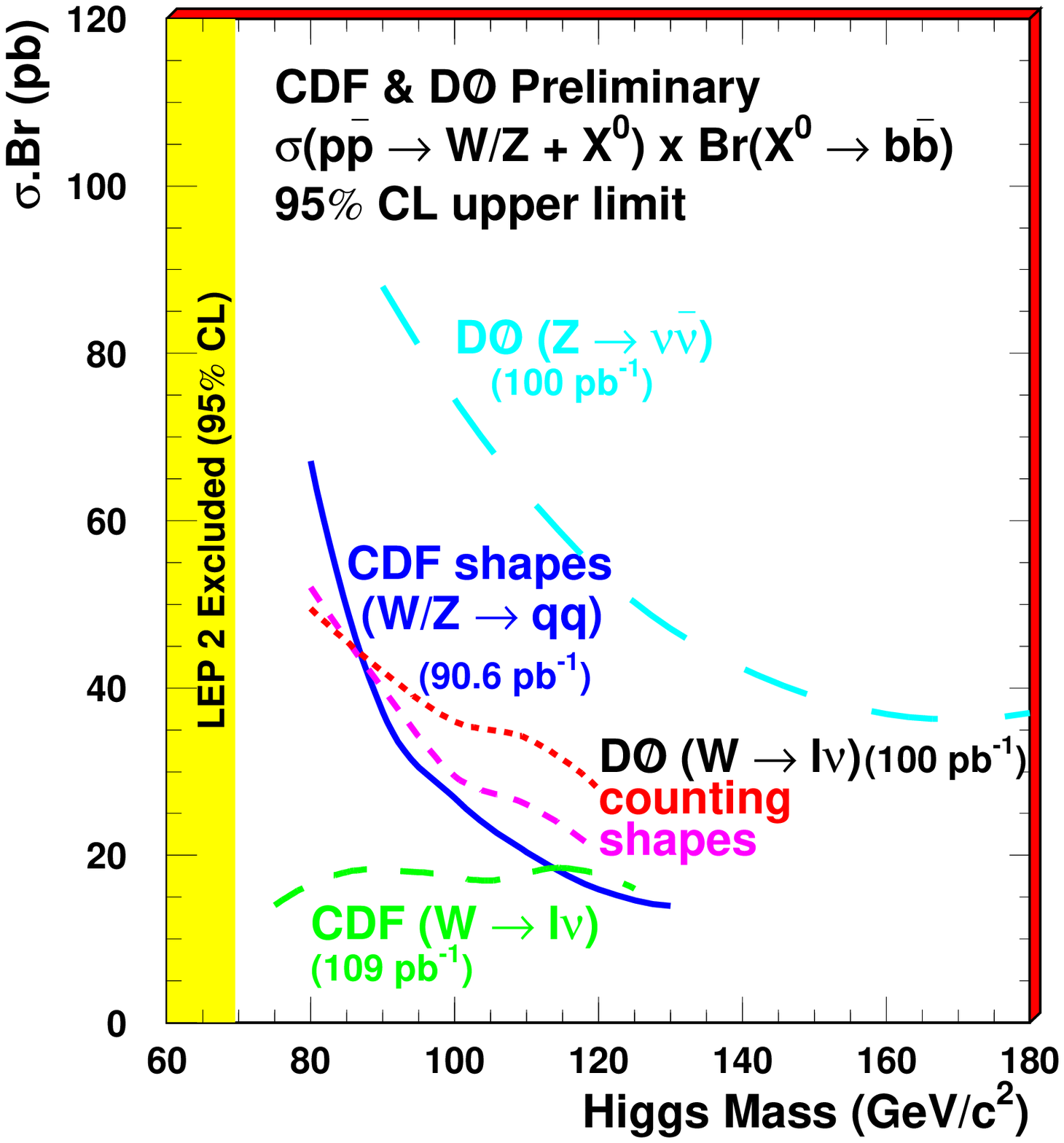}
\vspace*{1.5cm}
   \caption{Summary of associated Higgs
production limits with $H\rightarrow
b{\bar b}$ from the CDF and D$\O$
collaborations.}
   \label{Extra Higgs Limits}
\end{floatingfigure}
%\end{figure}

Both CDF and D$\O$ have used the $\ell\nu b{\bar b}$ decay channel to search for
and set limits on associated Higgs production ($VH$ where $V$ is a $W$ or a
$Z^0$) with 
$H\rightarrow b{\bar b}$~\cite{Higgs}.  These results~\cite{Higgs Leptonic}
are displayed along with 
results using the hadronic decays of the $W$ boson from
CDF 
and the  $Z^0 \rightarrow \nu{\bar \nu}$ results from 
D$\O$~\cite{D0 Z Higgs} in 
Figure~\ref{Extra Higgs Limits}.

\vspace{2.5cm}

% ***************************************************************
\section{Conclusions}

The CDF collaboration
continues to search for new phenomena at the Fermilab Tevatron and
has set limits on a number of Supersymmetric,
Technicolor and Higgs models. The CDF 
$\eeggmett$ candidate 
event has now been studied systematically.  However, one of the 
electron candidates has an ambiguous interpretation and 
the origin of the event remains a mystery. While 
many searches for
hints of its origin are complete (for some interpretations), 
many are still underway. 
With the Run II upgrades providing better acceptance and many times the
data the future is very promising.

\section{Acknowledgments}

The author would like to thank Henry Frisch, Ray Culbertson, Peter Wilson, Sarah
Eno, John Hobbs, John Womersley, Bryan Laurer, Takanobu Handa, Juan Valls,
Leslie Groer, Stephen Mrenna, Dave Hedin, Chris Kolda, Gordy Kane 
and Jianming Qian for their work
on the analyses presented.  

%
% Here we do the References
%


\clearpage

\begin{thebibliography}{99}

\bibitem{CDF Detector}  F. Abe, {\it et al.}, (CDF), Nucl. Instrum. Meth.
{\bf{271A}}.

%\bibitem{D0 Detector} S. Abachi, {\it et al.}, (D$\O$), Nucl. Instrum. Meth.
%{\bf{338A},}
%185 (1994) and references therein.

\bibitem{Higgs} 
A. Stange, W. Marciano, and S. WIllenbrock,
\Journal{\PRD}{49}{1354}{1994} and references therein.

\bibitem{Gravitino Reference}
a)~P.~Fayet, {\it Phys Rep} {\bf 105} 21 (1984);
b)~M.~Dine, A.~Nelson, Y.~Nir and Y.~Shirman, \Journal{\PRD}{53}{2658}{1996};
c)~S.~Dimopoulos, M.~Dine, S.~Raby and S.~Thomas, 
\Journal{\PRL}{76}{3494}{1996};
d)~S.~Dimopoulos, S.~Thomas and J.~Wells, \Journal{\PRD}{54}{3283}{1996};
%Sept 1st 1996
%Experimental consequences of a minimal messenger model for supersymmetry
%breaking.
e)~K.~Babu, C.~Kolda and F.~Wilczek, \Journal{\PRL}{77}{3070}{1996};
% hep-ph/9605408. Oct 7, 1996
%
% A supergravity explanation of the cdf e e gamma gamma event
f)~J.~Lopez and D.~Nanopoulos,  {\it Mod. Phys. Lett.} A{\bf 10}, 2473 (1996);
%\mbox{hep-ph/9607220;}
%Experimental consequences of no scale supergravity in light of the cdf e e
%gamma gamma event.
%ibid. hep-ph/9608275; see also 
%Flipped No-scale Supergravity: a Synopsis
%hep-ph/9701264.
%
%SEARCH FOR SUPERSYMMETRY WITH A LIGHT GRAVITINO AT THE 
%FERMILAB TEVATRON AND CERN LEP COLLIDERS
g)~S.~Ambrosanio, G.~Kane, G.~Kribs, 
S.~Martin and S.~Mrenna, \Journal{\PRD}{54}{5395}{1996}.
%hep-ph/9605398 

\bibitem{Higgsino LSP}
a)~H.~Haber, G.~Kane  and M.~Quiros, 
{\it Phys Letters} B {\bf 160}, 297 (1985); 
b)~S.~Ambrosanio, G.~Kane, G.~Kribs, S.~Martin and S.~Mrenna, 
\Journal{\PRL}{76}{3498}{1996};
% \mbox{hep-ph/9602239.}
c)~S.~Ambrosanio, G.~Kane, G.~Kribs, S.~Martin and
S.~Mrenna, \Journal{\PRD}{55}{1372}{1997}.
%hep-ph/9607414.%
%
%
\bibitem{Non-SUSY}
% A non supersymmetric interpretation of the CDF eeggmet event
See for example 
a)~G.~Bhattacharyya and R.~Mohapatra,  \Journal{\PRD}{54}{4204}{1996};
%\mbox{hep-ph/9606408.}
% An E(6) interpretation of an e+ e- gamma gamma missing e(t) event.
b)~J.~Rosner, \Journal{\PRD}{55}{3143}{1997};
%CERN-TH-96-209, Submitted to {\it Phys. Rev. D}.
% hep-ph/9607467;
c)~K.~Lane, {\it Phys Letters} B {\bf 357}, 624 (1995).
%e-Print Archive: hep-ph/9507289


\bibitem{GG PRL}  \mbox{F. Abe \etal}, \PrePrintE{9801019}. Submitted to Physics
Review Letters.

\bibitem{Techni Omega}  E.~Eichten, K.~Lane and J.~Womersley, 
\Journal{\PLB}{405}{305}{1997}

    \bibitem{Techni Rho}  
%E. Eichten and K. Lane {\it Two Scale Technicolor}  {\it Physics Letters} 
%{\bf 99B} (1980).
%K. Lane and M.V. Ramana 
%{\it Walking Technicolor signatures at Hadron Colliders} 
%\Journal{\PRD}{44}{2678}{1991}
%E. Eichten and K. Lane {\it Low-Scale Technicolor at the
%Tevatron} FERMILAB-PUB-96/075-T, BUHEP-96-9, hep-ph/9607213; to appear in
%Physics Letters B.
E. Eichten and K. Lane 
%{\it Electroweak and Flavor Dynamics at Hadron Colliders--I} 
\PrePrintP{9609297} and references therein. 
%; to appear in the proceedings of the 1996
%DPF/DPB Summer Study on New Directions for High Energy Physics (Snowmass 96).


\bibitem{D0 Bosonic Higgs} B.~Abbott \etal, 
%{\it Search for High Mass Photon Pairs in $p{\bar p}
%\rightarrow \gamma\gamma jj$ Events at \mbox{$\sqrt{s} = 1.8$~TeV}}, 
FERMILAB-CONF-97-325-E, Oct 1997. 



\bibitem{OPAL Bosonic Higgs} K. Ackerstaff \etal, Eur. Phys. J. {\bf C} 1, 31
(1998).

\bibitem{D0 GG Limits} 
S.~Abachi \etal,
\Journal{\PRL}{78}{2070}{1997} and B.~Abbott \etal, 
\Journal{\PRL}{80}{442}{1998}

\bibitem{LEP GG Limits}
ALEPH  Collaboration, {\it Phys. Lett. B} {\bf 420} 127 (1998),
%accepted by {\it Phys. Lett. B}, %CERN-PPE-97-122, \PrePrintE{9710009},
DELPHI Collaboration, Eur. Phys. J. C {\bf 1}, 1 (1998),
L3     Collaboration, {\it Phys. Lett. B} {\bf 415} 299 (1997),
% CERN-PPE-97-076,
OPAL   Collaboration, Submitted to Eur. Phys. J. C., \PrePrintE{9801024}.
%DELPHI Collaboration, submitted to {\it Zeit. f. Physik C}, CERN-PPE-97-107,
%OPAL   Collaboration, \Journal{\PLB}{391}{210}{1997}.

%    \bibitem{top} F. Abe et al., \Journal{\PRD}{50}{2966}{1994}

\bibitem{Future PRD} D. Toback, Searches for 
New Physics in Diphoton Events in $p{\bar p}$ collisions at
\mbox{$\roots= 1.8$~TeV},
Ph.D. thesis, University of Chicago, 1997; \mbox{F. Abe \etal},
to be submitted to Phys Rev D.


\bibitem{Higgs Leptonic} F.~Abe \etal, \Journal{\PRL}{79}{3819}{1997}, 
P.~Tamburello, 
%{\it Search for Dijet Resonances Produced in Association 
%with $W$ bosons at D$\O$,} 
FERMILAB-CONF-96-389-EC, Oct 1996.

%\bibitem{Higgs Hadronic} This is the hadronic Higgs search reference.

\bibitem{D0 Z Higgs}  B.~Abbott \etal,
%{\it Search for $ZX \rightarrow \nu{\bar \nu} b{\bar b}$
%Events in the D$\O$ Detector, }
FERMILAB-CONF-97-358-E, Oct 1997. 







%\bibitem{Higgs2} K. Lane, \PrePrint{9507289}. Did this become a hep-ph?


%    \bibitem{Altarelli} G. Altarelli, B. Mele and M. Ruiz-Altaba
%{\it Searching for new Heavy Vector Bosons in $P{\bar P}$ colliders}
% Z. Phys. C45, P109-P121 (1989)


%    \bibitem{Other Models} We note that these  
%methods are applicable for 
%searches  such as   $W'\rightarrow tb,  W'\rightarrow  WH,  H\rightarrow WW$ and
%associated Higgs ($WH$) production.

%    \bibitem{Higgs Hunters Guide} J. Gunion, H. Haber, G. Kane and S. Dawson,
%{\it The Higgs Hunter's Guide}. Frontiers in Physics series. 1990. 

%    \bibitem{Moriond} T. Kamon (CDF Collaboration), Fermilab-Conf-96/106-E,
%Proceedings of the XXXIst 
%Rencontres de Moriond (QCD), Les Arcs, Savoie, France,
%March 23-30, 1996.

%    \bibitem{Ramond} P. Ramond, {\it Ann. Rev. Nucl. Part. Sci.} {\bf 33} (1983)
%31.

%    \bibitem{Z Not  Hadronic}  Note that we are not requiring the jets to be 
%hadronic decays from quarks. Specifically, $Z^0 \rightarrow e^+e^-$ and $Z^0
%\rightarrow \tau^+\tau^-$ decays are included in this search by virtue of the
%fact that we are simply searching for two energy clusters in the detector.

%    \bibitem{Masses} The jet-jet and W+jet-jet invariant 
%masses are created using the
%electron, \Met, and the two jets with the highest E$_{\rm T}$ in the events with
%$|\eta|<2.0$. 
%We define the jet-jet mass to be 
%$m = \sqrt{(E_1 + E_2)^2 - ({\vec P_1} + {\vec P_2})^2}$. The 
%jet-jet mass resolution is roughly 10$\%$.
%To reconstruct the
%W+jet-jet mass we need the longitudinal component of the neutrino momentum
%($P_{\rm z}^{\nu}$) since it cannot be measured, as particles 
%exiting the detector with large $|\eta|$ can carry large high 
%longitudinal momenta. If we assume
%the mass of the electron-neutrino system to be equal to W 
%boson mass\cite{W+jet}
%(taken to
%be 80 GeV/c$^2$) then $P_{\rm z}^{\nu}$ is restricted to two possible values. 
%When the transverse mass is greater than 80 GeV/c$^2$, the two 
%an $P_{\rm z}^{\nu}$ solutions have 
%imaginary components. In such events the constraint is made to the value of the
%transverse mass instead of to 80 GeV/c$^2$ which gives two identical solutions.
%Once we have the neutrino solution(s), we find the invariant mass of the 
%W+jet-jet system by using the two  jets. We choose the  solution with the 
%lower invariant mass since studies have shown that it does a better job of 
%giving back the 
%correct $W'$ mass and width. We also use the same procedure on the data as for
%the  background and the signal
%Monte Carlo samples to reduce any  other systematic problems introduced by this
%method. 



%    \bibitem{W+jet} F. Abe {\it et al.,} Phys. Rev. Lett. {\bf 73} 2296 (1994) 

\end{thebibliography}
\end{document}